\documentclass[a4paper,11pt]{article}
\usepackage{pos}

\title{Lee-Yang edge singularities in lattice QCD : A systematic study of singularities in the complex $\mu_B$ plane using rational approximations.}
\ShortTitle{Probing complex singularities in QCD à la Padè}

\author*[a]{Simran Singh}
\manuallySeparateAuthors
\author[a]{, Petros Dimopoulos,}
\author[b]{Lorenzo Dini,}
\author[a]{Francesco Di Renzo,}
\author[b]{Jishnu Goswami,}
\author[b]{Guido Nicotra,}
\author[b]{Christian Schmidt,}
\author[a]{Kevin Zambello,}
\author[b]{and Felix Ziesche}
\author{ (Bielefeld-Parma Collaboration)}

\affiliation[a]{Dipartimento di Scienze Matematiche, Fisiche e Informatiche, Università di Parma and INFN, \\Gruppo Collegato di Parma
  I-43100, Parma, Italy}
  
\affiliation[b]{Fakult$\ddot{a}$t f$\ddot{u}$r Physik, Universit$\ddot{a}$t Bielefeld,\\
D-33615 Bielefeld, Germany}

\emailAdd{simran.singh@unipr.it}
\emailAdd{petros.dimopoulos@unipr.it}
\emailAdd{francesco.direnzo@unipr.it}
\emailAdd{kevin.zambello@studenti.unipr.it}
\emailAdd{lorenzo@physik.uni-bielefeld.de}
\emailAdd{schmidt@physik.uni-bielefeld.de}
\emailAdd{jishnu@physik.uni-bielefeld.de}
\emailAdd{gnicotra@physik.uni-bielefeld.de}
\emailAdd{fziesche@physik.uni-bielefeld.de}

\abstract{A new approach is presented to explore the singularity structure of lattice QCD at imaginary chemical potential. Our method can be seen as a combination of the Taylor expansion and analytic continuation approaches. Its novelty lies in using rational (Padé) approximants for studying Lee Yang edge singularities. The motivation for using rational approximants will be exhibited. We will provide some confidence in our approach based on numerical experiments performed on well-
motivated “toy models”. Our focus lies in identifying singularities of the net-baryon number density in the complex $\mu_B$ plane. To this end we have found signatures of the Roberge-Weiss critical point(and Chiral singularities - subject to some caveats). In this contribution we will discuss the setup, simulation parameters and results obtained for 2+1 $flavor$ QCD in the complex $\mu_B/T$ plane.}

\FullConference{%
 The 38th International Symposium on Lattice Field Theory, LATTICE2021
  26th-30th July, 2021
  Zoom/Gather@Massachusetts Institute of Technology
}

\begin{document}
\maketitle

\section{Introduction}
\noindent One of the major open problems in high energy physics today is understanding the phase diagram of QCD. More specifically, the goal is to look for the QCD critical end point (if it exists). On the one hand, based on lattice QCD simulations at zero chemical potential a crossover is observed at T$_{pc}\sim $156 MeV \cite{HOTQCD}. On the other hand, a first order phase transition is expected at low temperature and high baryon number density. This prompts us to expect a critical end point somewhere in between these two transitions, in the QCD phase diagram. 

\noindent In principle direct simulations with lattice QCD should have been our best tool to look for such a critical point, but in practice this is not true due to the infamous \emph{sign problem}. At finite baryon densities ($\mu_B \neq 0$), due to the Grassmann nature of the fermion action we end up with an effective action that is complex. This complex action leads to a failure of interpreting the partition function as a "weighted sum" of the exponential of the action, as the "weight" is no longer positive definite. This leads to \emph{a numerical problem} which prevents direct Monte Carlo(MC) simulations of lattice QCD at finite $\mu_B$ - hence we look at \emph{indirect} methods.   
\section{Overview of current methods}
The current state of the art methods and their respective drawbacks include:
\begin{itemize}
\item[I] \emph{Taylor expansion} about $\mu_B=0$ (\cite{Allton:2002ns}). Since there is no sign problem at zero chemical potential it is possible to perform MC simulations directly using the full partition function and also evaluate higher derivatives of $\log{\mathcal{Z}_{GC}}$ (higher cumulants). The main drawback of this approach is that simulations get harder for computing higher derivatives and until date only a few Taylor coefficients are known precisely.
\item[II] \emph{Imaginary $\mu_B$} simulations (\cite{DEFORCRAND2002290}, \cite{PhysRevD.67.014505}). In this setup there is no sign problem, so in principle simulations can be performed and combined with Taylor expansions, continued back for real values of $\mu_B$.The problem encountered here is that the partition function inherits a remnant of the \emph{centre symmetry} of the Yang Mills field which causes certain non-analyticities (Roberge-Weiss phase transition \cite{Roberge:1986mm}) which prevent analytic continuation of thermodynamic variables beyond a certain value of the chemical potential.
\end{itemize} 
\noindent \emph{Our approach can be thought of as a combination of the above mentioned Taylor expansion and Imaginary $\mu_B$ approaches.}\footnote{In the context of QCD, the idea has been put forward before by Maria Paola Lombardo \cite{Lombardo:2005ks} and Rajiv Gavai \cite{RGavai2008} independently. Another work using Padé type approximations had appeared recently \cite{Pasztor} aimed at studying the crossover line - but to our knowledge this is the first attempt to use Padé to directly probe singularities in the phase diagram.} The idea is to use (low order) Taylor coefficients at $\mu_B = 0$ and  at imaginary values to build a rational function \emph{à la} Padé. This is much the same in spirit as the approach used in \cite{DiRenzo:2020cgp}. The main purpose of using rational functions is \emph{the recovery of singularities of a function}, which is to be expected if we are looking for phase transitions in our system.
\section{Our approach : Padè approximants}
\begin{itemize}
\item[o] \emph{Single-point Padé} : This is the original method of constructing a rational function out of Taylor coefficients and requires the knowledge of coeffcients at a single point : 
\begin{equation*}
f(x) = \sum\limits_{i=0}^L \, c_i \, x^i + \mathcal{O}(x^{L+1}) \approx R^{m}_{n}(x) = \frac{P_m(x)}{1 + Q_n(x)} =\frac{\sum\limits_{i=0}^m \, a_i \, x^i}{1 + \sum\limits_{j=1}^n \, b_j \, x^j} 
\end{equation*}
\noindent The obvious drawback of this method, for us, is the knowledge of a high number of coefficients at a single point - which we don't have for QCD at any point in phase space. 

\item[o] \emph{Multi-point Padé} : A natural extension of the above approach is when we have lower order Taylor coefficients but at multiple points. We can then modify the above equation to solve a system of linear equations:
\begin{equation*}
\begin{split}
P_m(x_i) - f(x_i)Q_n(x_i) &= f(x_i) \\ 
P_m'(x_i) - f'(x_i)Q_n(x_i) - f(x_i)Q_n'(x_i) &= f'(x_i) \\ 
 & \hdots \\
\end{split}
\end{equation*}

\noindent Within our collaboration \cite{NEwColP} other methods of obtaining the multi-point Padé approximation exist, but here we mainly focus on the Linear Solver approach. One of these is a \emph{generalised $\chi^2$} approach, in which the difference of the rational ansatz with the known Taylor coefficients of the function are minimized with respect to the error bars on the known coefficients. 
\end{itemize}
\subsection{Observations based on numerical experiments : }
\begin{figure}[h!] 
        \includegraphics[scale=0.335]{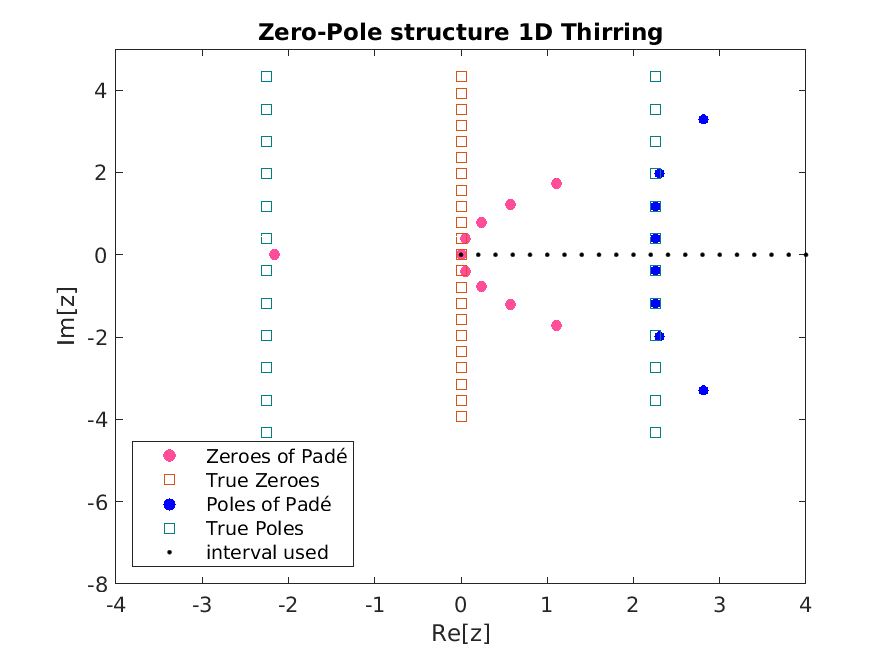}
        \includegraphics[scale=0.335]{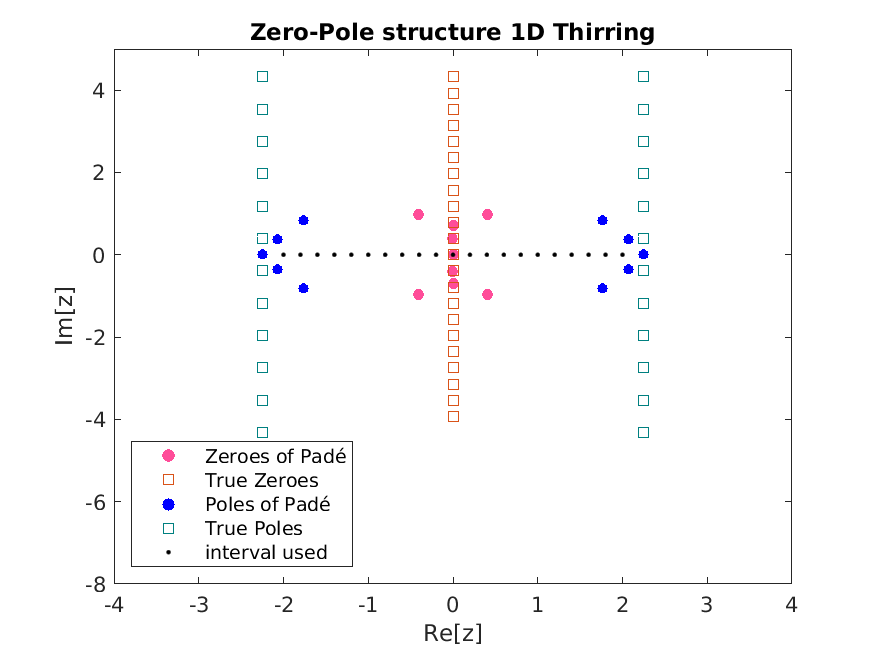}
        \includegraphics[scale=0.335]{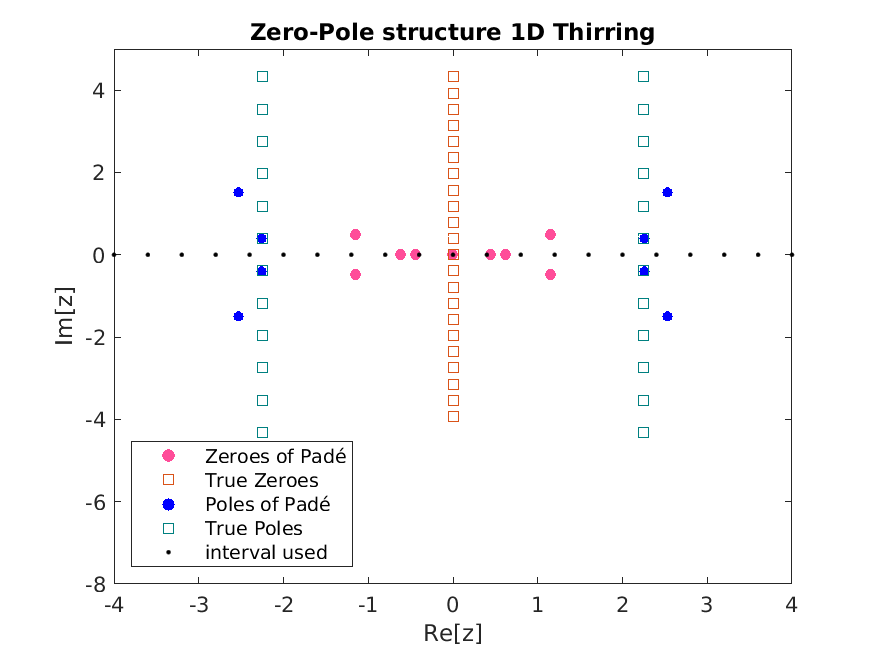}
        \caption{\scriptsize{Interval sensitivity of Multi-Point Padé : (Left) With $Re[z]\,\in [0,4]$ which is centered around $Re[z]=2$, sensitivity is enchanced for the analytic poles at $Re[z] \sim 2.25$, (Centre) Sensitivity is enhanced for the zeroes at $Re[z]=0$, (Right) Signature for the closest poles zeroes throughout the interval are present}}
        \label{fig:IntervalSen}
\end{figure}

\begin{figure}[h!] 
        \centering
        \includegraphics[scale=0.35]{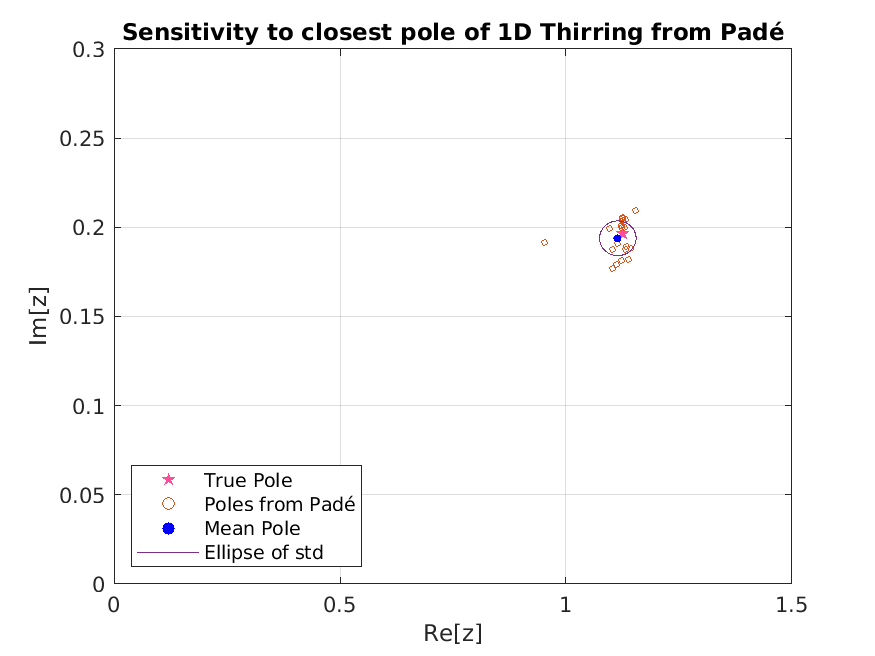}
        \hfill
        \includegraphics[scale=0.35]{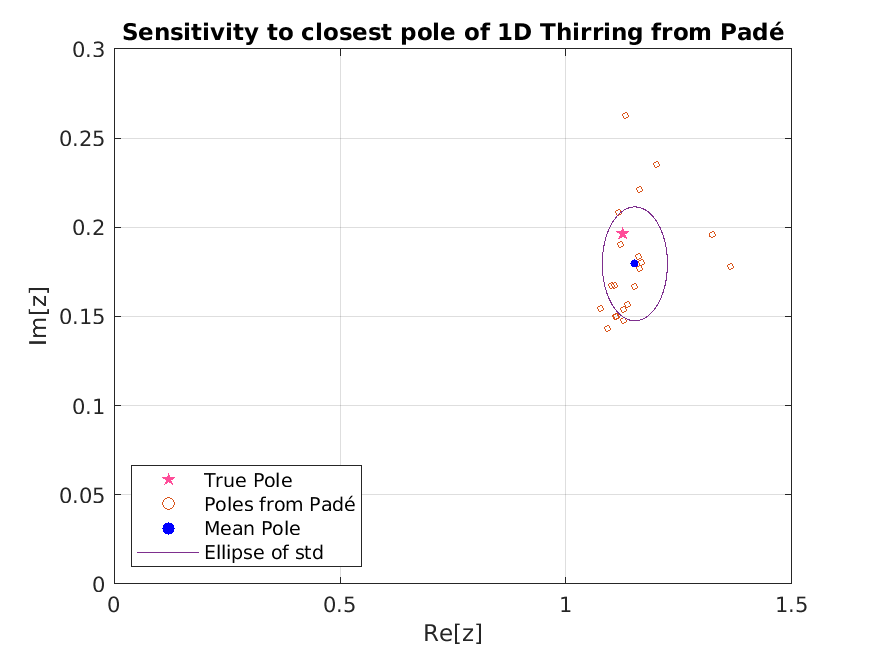}
        \caption{\scriptsize{1D Thirring model : Sensitivity to the closest pole even in the presence of errors! (Left) : 1\% noise on values and 10\% noise on 1st derivative and (Right) : 5\% noise on values and 15\% noise on 1st derivative for [6/6] Padé}}
        \label{fig:closeip}
\end{figure}

\noindent While a large amount of literature (\cite{baker1975},\cite{NUTTALL1970147}) exists on single point Padé approximations - ranging from existence, uniqueness and convergence theorems - this is not true in general for the multi-point Padé approach. However, in order to understand the scope of validity of this method, multiple numerical experiments were performed on known functions, also using noisy data . A few of the results of importance are listed below:
\begin{itemize}
\item[o]\textit{Poles and other singularities obtained from an MP is interval sensitive} Fig. \ref{fig:IntervalSen}. This means that, even though the closest singularity is usually observed with this approach - we maybe more (or less) sensitive to certain zeroes or poles depending on where we sample the Taylor coefficients.
\item[o] Padé theory dictates that a \textit{genuine pole of the function will remain stable when changing order of the Padé.} While this is true for clean data (functions without noise) - the picture changes when we introduce noise \cite{baker1975}.
\item[o] In the presence of noise we are still sensitive to the closest singularity w.r.t. the axis we expand about. \textit{The only catch being that now the pole moves about roughly in an ellipse the size of the magnitude of error introduced} (see Fig. \ref{fig:closeip}).
\end{itemize}

\subsection{A short note on spurious poles}
\noindent A common criticism that Padé-type approximations receive are based on encountering \emph{spurious poles}. These can occur even for clean data - but are not as harmful as they are perceived to be. Based on our numerical exercises, we have found that spurious poles usually occur in the following three forms:
{\tt (i)} \textit{Exactly canceling pairs of zeros and poles of the approximation}, these can occur even for noise free data. This usually results from demanding a very high order of the rational approximation from the given data. So if we define our rational approximation to be strictly co-prime with respect to zeros and poles - these should be removed. {\tt (ii)} \textit{Isolated poles can sometimes occur} when we choose particular datasets. But even these are not that harmful, since they only show up for certain orders of the approximation - i.e \textit{they are never stable}. In many cases some poles show up which move away to infinity when we systematically increase the order of the Padé. {\tt (iii)} When simulating data with noise, we mostly get \textit{nearly canceling zero pole pairs}. These are \textit{slightly different from the exactly canceling} ones we mentioned before, because they occur even at low order approximation. On average we miss exact cancellation by an amount proportional to the magnitude of noise introduced.

\section{Overview of QCD at imaginary $\mu$}
\noindent Pure gauge theory (SU(N) Yang Mills) has a $\mathbb{Z}_N$ centre symmetry at low temperatures and undergoes spontaneous symmetry breaking at high temperatures. This symmetry is explicity broken when fermions are added to the YM partition function (in the form of the chemical potential $\mu_B$). However, at purely imaginary values of the chemical potential, the QCD partition function inherits a remnant of the Yang Mills centre symmetry. This manifests as $\mathcal{Z}_{GC}(\mu_B + i 2 \pi T) = \mathcal{Z}_{GC}(\mu_B)$, the partition function becomes periodic. In addition to this, the partition function also has a charge conjugation symmetry given by, $\mathcal{Z}_{GC}(\mu_B) = \mathcal{Z}_{GC}(-\mu_B)$.

\noindent Moreover, there is an expectation of a phase transition at all temperatures above the Roberge-Weiss critical end point $T_{RW}$ \cite{Roberge:1986mm} for QCD at imaginary $\mu_B$. The (imaginary) baryon number density ($\chi^1_B(T,\mu_B)$) is expected to become discontinuous for $T > T_{RW}$ \cite{Borsanyi2018}. Given the symmetries of the partition function, and the relation of $\chi^1_B$ with the partition function :
\begin{align}\label{chiEqn}
\chi^1_B(T, \hat{\mu}_B) = \frac{n_B(T,\hat{\mu}_B)}{T^3} = \frac{\partial ( p(T,\hat{\mu}_B)/T^4)}{\partial \hat{\mu}_B} = \frac{1}{{(VT)}^3 \mathcal{Z}} \, \frac{d \mathcal{Z}}{d \hat{\mu}_B}
\end{align}
$\chi^1_B$ is an \emph{odd and periodic} function of $\mu_B/T$. Hence, $\chi^1_B$ acts as our \emph{order parameter} to "hunt" for the \emph{Roberge Weiss critical point}.

\subsection{Details of simulation setup}
\noindent Baryon number density and its derivatives were obtained as a function of imaginary baryon chemical potential using highly improved staggered quarks (HISQ) on a $24^3 \times 4$ lattice for five different temperatures. The simulations were performed at physical quark mass ($m_s/m_l = 27$) and $\mu_B/3 = \mu_l = \mu_s$. For this particular lattice discretization and temporal extent, the Roberge-Weiss temperature is $T_{RW} = 201$ MeV \cite{JGowswami}. Simulations were also performed on N$_\tau = 6$ lattices but with the purpose of being sensitive to chiral singularities. Some further details can be found in the table below :
\begin{center}
\begin{figure*}[h!]
\centering
\scalebox{0.63}{
\begin{tabular}{|c|c|c|c|c|c| } 
\hline
\multicolumn{1}{|c|}{} & \multicolumn{5}{|c|}{$N^3_\sigma \times N_\tau$ = $24^3 \times 4$} \\
\hline
$T$ (MeV) &  160 & 167.38 & 176 & 186.26 & 201\\ 
\hline
\#confs & 5550 & 6000 (12K) & $\sim$ 2500 & 2500 & 5000 (11K) \\
\hline
\multicolumn{1}{|c|}{} & \multicolumn{5}{|c|}{$N^3_\sigma \times N_\tau$ = $36^3 \times 6$} \\
\hline
\multicolumn{1}{|c|}{$T$ (MeV)} & \multicolumn{2}{|c|}{125} & \multicolumn{3}{|c|}{145} \\
\hline
\multicolumn{1}{|c|}{\#confs} & \multicolumn{2}{|c|}{5280} & \multicolumn{3}{|c|}{8000} \\
\hline
\end{tabular}}
 \caption*{\scriptsize{Number of gauge configurations sampled for various temperatures. For scale setting see \cite{Bazavov2012}}}
\end{figure*}
\end{center}
\vspace{-1cm}
\begin{figure*}[h!] 
        \includegraphics[scale=0.25]{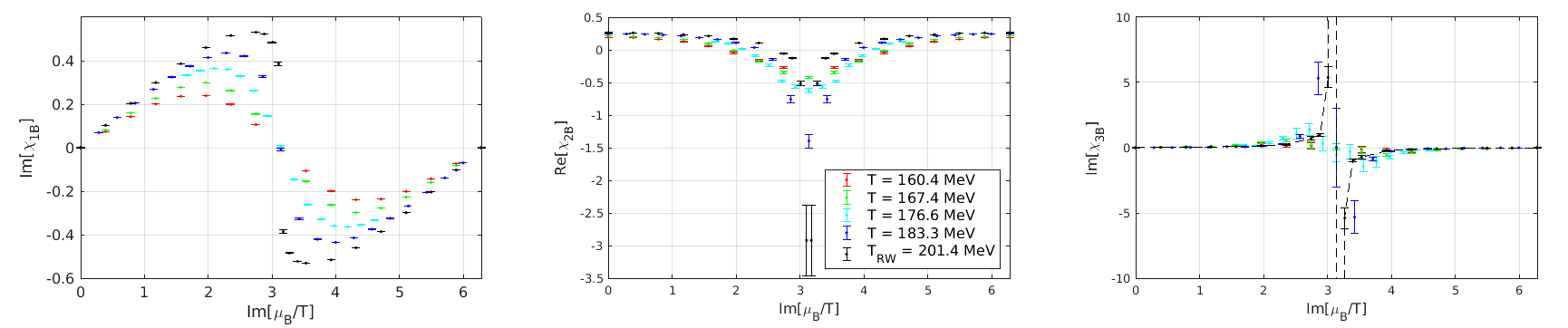}
        \caption{\scriptsize{Cumulants of the net baryon number fluctuations as function of Im[$\mu_B/T$] for the $24^3 \times 4$ lattices.}}
\end{figure*}
\section{Results}
\noindent Since we studied the baryon number density as our order parameter, the \emph{zeroes} of the partition function are accessible to us as \emph{poles} of the baryon density function (see Eq \ref{chiEqn}). Listed in Table \ref{tab:sings} are the values of the thermal singularities found for the N$_{\tau}=4$ case with error bars. 
\begin{table}[h!]
\scalebox{0.8}{
    \begin{tabular}{|c|c|c|c|c|c|c|c|c|}
    \hline
\multicolumn{1}{|c|}{$T$ (MeV)} & \multicolumn{2}{|c|}{ Method A} & \multicolumn{2}{|c|}{ Method B} & \multicolumn{2}{|c|}{ Method C*} & \multicolumn{2}{|c|}{ Method C} \\ \hline
 & $\mu_{LY}^{R}$ & $\mu_{LY}^{I}$ & $\mu_{LY}^{R}$ & $\mu_{LY}^{I}$ & $\mu_{LY}^{R}$ & $\mu_{LY}^{I}$ & $z^{R}$ & $z^{I}$\\ \hline  
201.4 & 0.11(11) & 3.142(10) & 0.077(45) & 3.133(15) & 0.0541(15) & 3.1294(63) & -0.9472(14) & -0.0116(60) \\
 186.3 & 0.48(14) & 3.118(54) & 0.53(13) & 3.112(66) & 0.397(51) & 3.127(34) & -0.672(34) & 0.010(21) \\
 176.6 & 1.03(10) & 3.112(72) & 1.022(80) & 3.18(12) & 1.040(94) & 3.115(65) & -0.353(33) & -0.010(20) \\
 167.4 & 1.82(11) & 3.125(79) & 1.79(13) & 3.164(95) & 1.694(55) & 3.12(13) & -0.184(12) & 0.004(22) \\
 160.4 & 2.097(90) & 3.147(11) & 2.14(12) & 3.150(70) & 2.07(76) & 3.14(24) & -0.126(70) & 0.000(14) \\ \hline
    \end{tabular}}
    \caption{\scriptsize{Method A : Linear Solver. Method B : $\chi^2$ fit approach. Method C : Linear solver in fugacity plane (discussed in \cite{NEwColP}).  (Note* : Mapped back values from fugacity plane. We are picking the value in first quadrant given the symmetries of the partition function)}}
    \label{tab:sings}
\end{table}

\subsection{Singularity structure in complex $\mu_B/T$ plane}
\noindent Displayed below are results shown for the central data of N$_{\tau}=4$ data. It is easy to see that the only "un-cancelled" zeroes and poles are the expected ones. Although only one slice of the structure is shown per temperature - it has been checked that the relevant poles and zeroes are stable within error bars.

\begin{figure*}[h] 
        \includegraphics[scale=0.335]{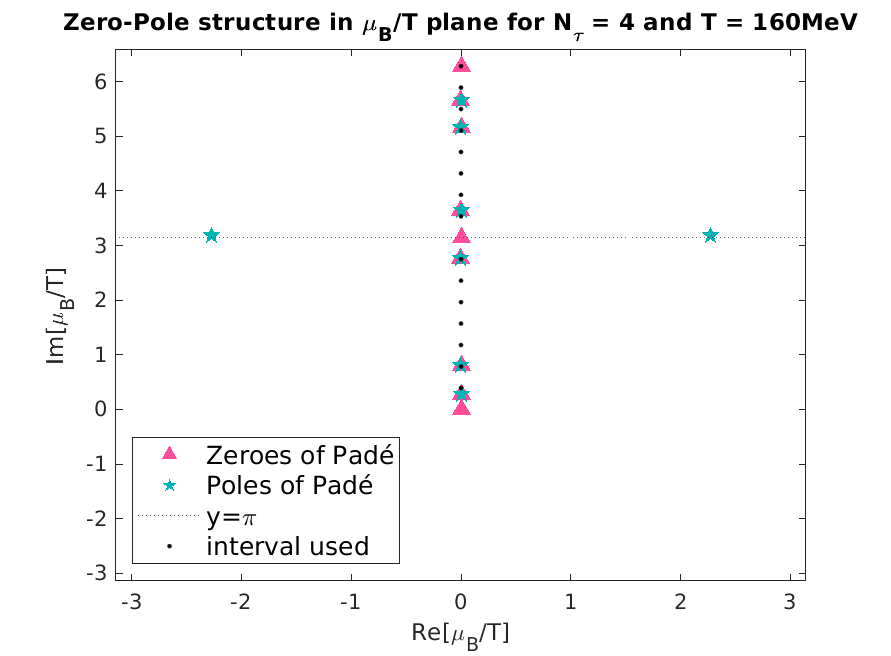}
        \includegraphics[scale=0.335]{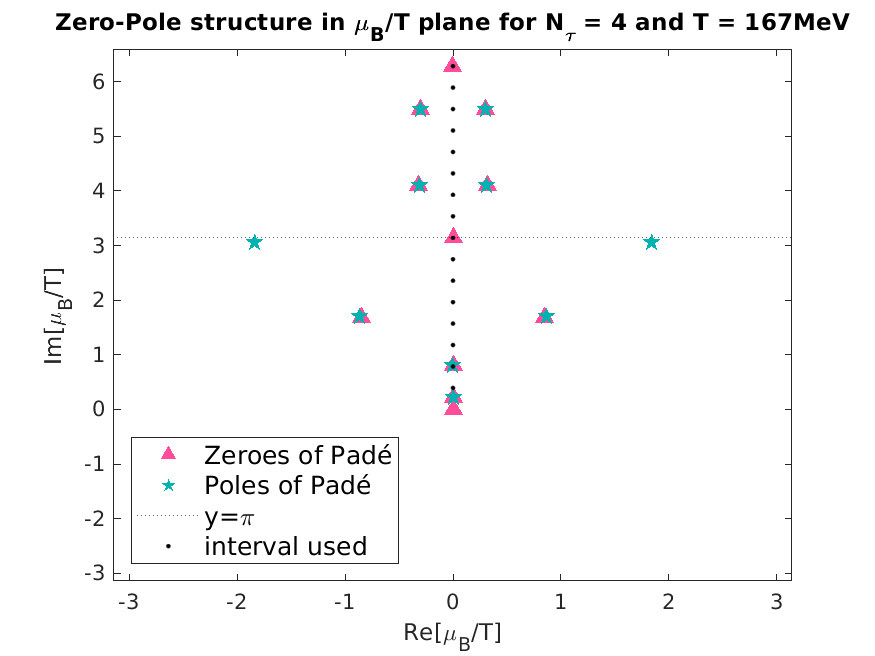}
        \includegraphics[scale=0.335]{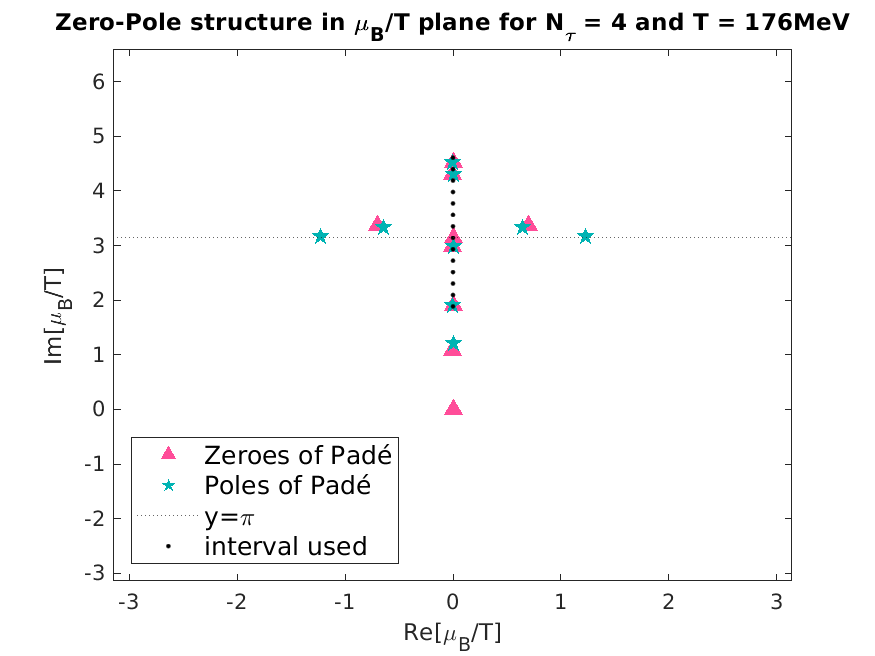}
        \includegraphics[scale=0.335]{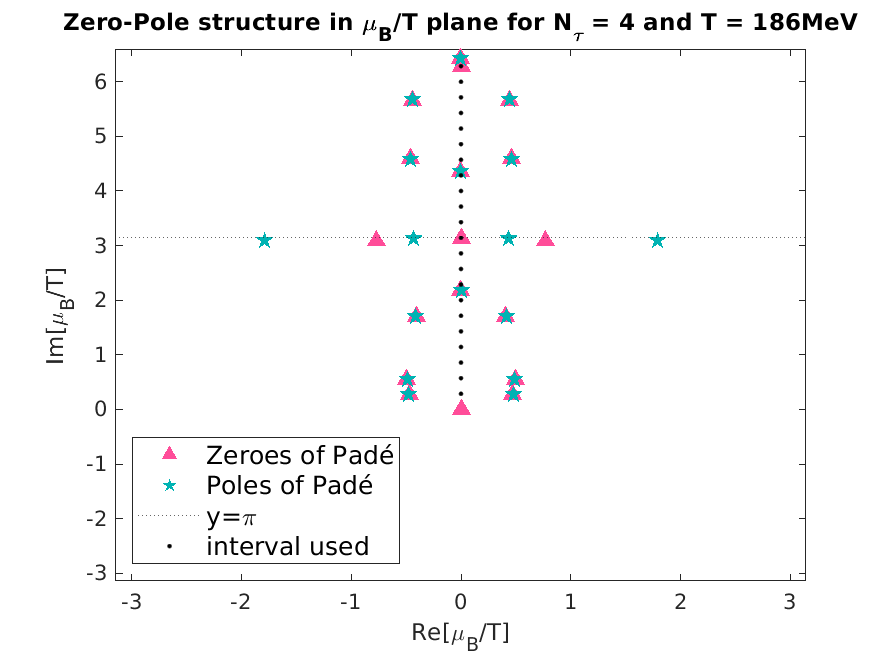}
        \includegraphics[scale=0.335]{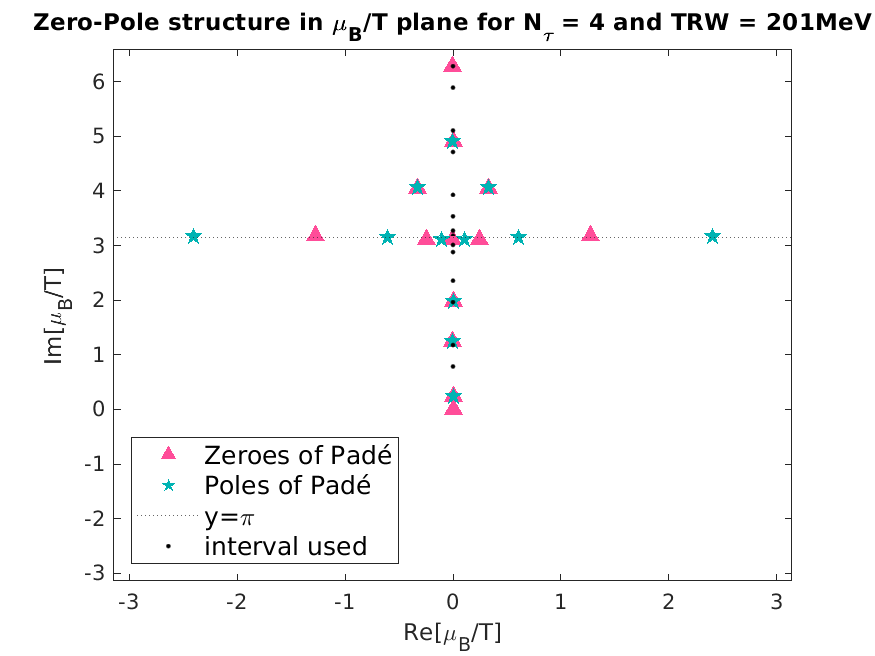}        
        \caption{\scriptsize{Zero-Pole structure of the approximation in the "original" $\mu_B/T$ plane from the lowest (top left) to the highest (bottom right) temperatures.}}
        \label{fig:SingularMuT}
\end{figure*}

\subsection{Rational approximations and Analytic continuation}
\noindent Before concluding anything regarding the singularity structure, we must convince the reader that the approximation has the correct behaviour. Shown below in Fig. \ref{fig:Ratanalytic} are a few functional forms of the approximations obtained and their analytic continuation. Also shown there, are the free energy plots for three distinct temperatures obtained from integrating the rational approximations.
\begin{figure*}[h!] 
        \includegraphics[scale=0.28]{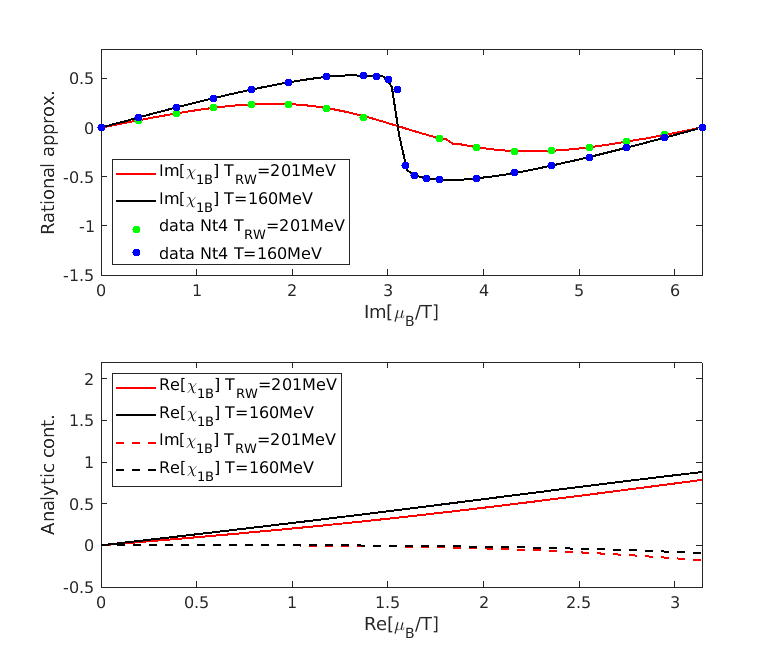}
        \hfill
        \includegraphics[scale=0.45]{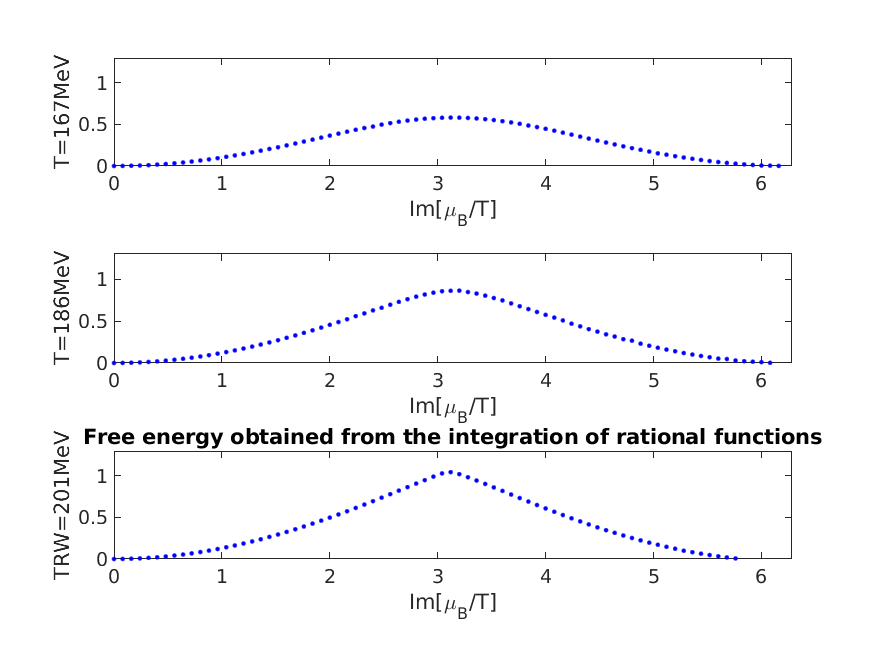}    
        \caption{\scriptsize{ (Left) Rational approximation and Analytic continuation for the lowest and highest temperatures. (Right) Free energy plots for three temperatures shown.}}
        \label{fig:Ratanalytic}
\end{figure*}

\subsection{Bootstrapping the errors}
\noindent It is not enough to conclude anything about the Lee Yang edge singularity structure without understanding how the error in the data obtained from simulating the baryon number density propagates into the poles. For this reason we performed a "bootstrap" procedure where we varied the input coefficients going into the Padé solver about the central values with standard deviation being the error bars on the respective data. This results in the following error bars on the poles obtained.

\begin{figure*}[h!] 
\centering
        \includegraphics[scale=0.4]{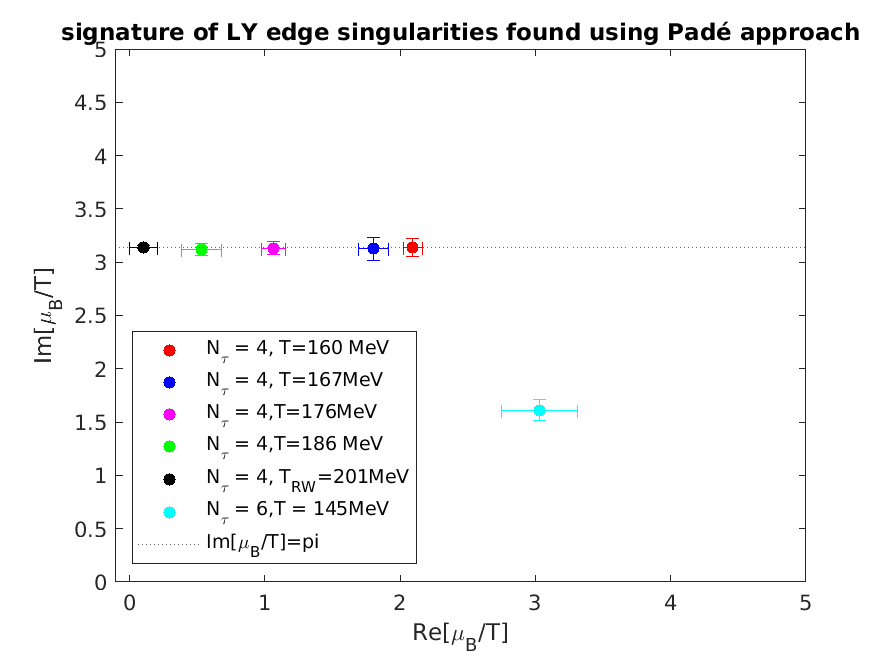}    
        \caption{\scriptsize{ Lee Yang edge singularities obtained from the bootstrap procedure. More on the cyan point from the N$_{\tau}$=6 data is discussed in \cite{NEwColP}}}
        \label{fig:polesEr}
\end{figure*}

\noindent Based on a recent paper \cite{Basar2021} and an older one \cite{Skokov:2010uc}, we decided to test our results using conformal maps\footnote{This work was done after the conference hence will only be outlined here}. The motivation of doing this was two-fold : One being increasing sensitivity of the series coefficients to certain singularities which can help us minimise the sensitivity to "spurious singularities" if there are any. Another reason to consider conformal maps (or even other "scaling" maps) is to reduce the ill-conditioning of the linear system (if any). 

\section{Conclusions and Outlook}
\noindent Through this contribution we have tried to show the power of using Padé type approximants to detect non-analyticities (poles, branch cuts etc.) in thermodynamic quantities of QCD at imaginary chemical potential. These non-analyticities can be related to the zeroes of the partition function via Lee-Yang edge singularities (LYEs). We have broadly outlined the scope of validity of such rational approximations and shown through numerical experiments how to distinguish between \emph{genuine} and \emph{spurious} singularities.  

\noindent While the goal of this contribution was to motivate the study of lattice QCD data using Padé approximations, the more important result of this work was from the scaling analysis of the Roberge-Weiss point in the N$_{\tau}=4$ lattice. The scaling of the LYE singularities follows the expected one for non-continuum extrapolated data, i.e, of the 2nd order Z(2) critical point (This will be explained in detail in \cite{GuidoProc} and references are cited therein). Simulations were also performed on the N$_{\tau}=6$ lattices with the goal of detecting Chiral singularities. In this case we are sensitive to the signature of a singularity which is consistent with Chiral scaling but more statistics are needed to be certain.

\noindent The very next steps of the project are to determine the Roberge-Weiss LYEs in the N$_{\tau}=6$ lattices. For this work, computing time within PRACE 2021 was granted and the work has started. 

\acknowledgments
\noindent This work was supported by {\tt (i)} the European Union’s Horizon 2020 research and innovation program under the Marie Skłodowska-Curie grant agreement No H2020-MSCAITN-2018-813942 (EuroPLEx), {\tt (ii)} The Deutsche Forschungsgemeinschaft (DFG, German Research Foundation) - Project number 315477589-TRR 211 and {\tt (iii)} I.N.F.N. under the research project {\em i.s.} QCDLAT.

\noindent This research used computing resources made available through: {\tt (i)} Gauss Centre for Supercomputing on the Juwels GPU nodes at the  Jülich Supercomputing Centre {\tt (ii)} Bielefeld University on the Bielefeld GPU-Cluster {\tt (iii)} CINECA on Marconi100 under both the I.N.F.N.-CINECA agreement and the ISCRA C program (HP10CWD9YA project) and {\tt (iv)} University of Parma on the UNIPR HPC facility.

\end{document}